\documentclass[acmsmall,manuscript,screen]{acmart}

\usepackage{tabularx}
\usepackage{verbatim}
\usepackage{booktabs}
\usepackage{longtable}
\usepackage{array}
\usepackage{multirow}
\usepackage{caption}
\usepackage{lscape}
\renewcommand{\arraystretch}{1.5}
\usepackage{ragged2e} 
\usepackage{ltablex}
\keepXColumns

\renewcommand{\arraystretch}{1.2}
\AtBeginDocument{%
  }

\setcopyright{none}
\copyrightyear{2024}
\acmYear{2024}
\acmDOI{}

\begin{document}

\title{A Survey on Methodological Approaches to Collaborative Embodiment in Virtual Reality}


\author{Hongyu Zhou}
\orcid{0009-0007-3278-2122}
\affiliation{%
  \institution{The University of Sydney}
  \department{School of Computer Science}
  \country{Australia}
}
\email{hzho4130@uni.sydney.edu.au}

\author{Yihao Dong}
\orcid{0009-0009-0719-3670}
\affiliation{%
  \institution{The University of Sydney}
  \department{School of Computer Science}
  \country{Australia}
}
\email{yihao.dong@sydney.edu.au}

\author{Masahiko Inami}
\orcid{0000-0002-8652-0730}
\affiliation{%
  \institution{The University of Tokyo}
  \department{Research Center for Advanced Science and Technology}
  \city{Tokyo}
  \country{Japan}
}
\email{drinami@star.rcast.u-tokyo.ac.jp}

\author{Zhanna Sarsenbayeva}
\orcid{0000-0002-1247-6036}
\affiliation{%
  \institution{The University of Sydney}
  \department{School of Computer Science}
  \country{Australia}
}
\email{zhanna.sarsenbayeva@sydney.edu.au}

\author{Anusha Withana}
\orcid{0000-0001-6587-1278}
\affiliation{%
  \institution{The University of Sydney}
  \department{School of Computer Science, Sydney Nano Institute}
  \country{Australia}
}
\email{anusha.withana@sydney.edu.au}




\renewcommand{\shortauthors}{H. Zhou et al.}

\begin{abstract}
  The application and implementation of collaborative embodiment in virtual reality (VR) are a critical aspect of the computer science landscape, aiming to enhance multi-user interaction and teamwork in immersive environments. A notable and enduring area of collaborative embodiment research focuses on approaches that enable multiple users to share control, interact, and investigate scenarios involving supernumerary arms in virtual spaces. In this survey, we will present an extensive overview of the methodologies employed in the past decade to enable collaboration in VR environments, particularly through embodiment. Using the PRISMA guidelines, we plan to analyze the study details from over 137 relevant research papers. Through this analysis, a critical assessment of the effectiveness of these methodologies will be conducted, highlighting current challenges and limitations in implementing collaborative embodiment in VR. Lastly, we discuss potential future research directions and opportunities for enhancing collaboration embodiment in virtual environments. 
\end{abstract}



\thanks{This paper is currently under review by ACM Computing Surveys (CSUR).}
\maketitle


\section{INTRODUCTION}
Virtual reality (VR) has become increasingly prominent in fields such as education~\cite{kavanagh2017systematic}, healthcare~\cite{rose2018immersion}, gaming~\cite{cruz2018virtual}, and remote collaboration~\cite{huang2023virtual}, providing immersive environments that improve interaction and user experience~\cite{zhou2024coplayingvr,zhou2025juggling}. A key element of these advancements is the concept of collaborative embodiment, which extends self-location, agency, body ownership, and the integration of virtual bodies from individual experiences to multi-user interactions~\cite{gonzalez2018avatar,hapuarachchi2022knowing}. By exploring and enhancing these aspects, collaborative embodiment contributes to greater immersion, interactivity, and realism, supporting more natural and intuitive collaborative user interactions. For example, in applications such as education~\cite{slater2017implicit}, gaming~\cite{zhou2024coplayingvr}, and remote collaboration~\cite{freeman2022working}, effective collaboration within shared virtual spaces is crucial for enhancing user engagement and interaction. 
As VR continues to evolve, developing methodologies for collaborative embodiment, where multiple users interact and work together seamlessly, is becoming increasingly important.

Although substantial progress has been made in understanding individual embodiment in VR, the collaborative aspects of embodiment, where multiple users engage within the same virtual space, have not been fully explored. While this area is crucial for enhancing human interaction in VR, the methodologies needed to effectively facilitate collaborative experiences have not yet undergone a comprehensive review. Collaboration within virtual spaces presents unique challenges, such as maintaining synchronization between users, balancing control among multiple participants, and preserving the sense of immersion during collective tasks~\cite{doroudian2023collaboration}. As VR applications expand across diverse fields like education~\cite{dunmoye2022exploratory}, gaming~\cite{zhou2024coplayingvr}, and remote collaboration~\cite{huang2023virtual}, addressing these challenges has become essential for creating more fluid and natural multi-user experiences. There is a growing need to investigate how embodiment techniques can support such collaboration, offering users not only individual immersion but also seamless interaction with others in shared virtual environments. This survey provides a comprehensive review of the methodologies developed to facilitate collaborative embodiment in virtual environments, where multiple users interact and work together seamlessly. It offers a much-needed overview of the methodological concerns and strategies employed to address issues such as synchronization, shared control, and immersive interaction in multi-user VR settings. Specifically, our article provides a systematic review of methodological approaches to enable collaborative embodiment.

This survey is motivated by the need to systematically understand three critical aspects of collaborative embodiment studies: the measures and metrics employed, the task designs used, and the collaboration methods explored. Analyzing these aspects is essential to identify methodological strengths, reveal limitations, and inform future study designs in collaborative virtual environments. By providing this systematic analysis, our goal is to offer future researchers clear guidance on evaluation standards, task choices, and collaborative mechanisms that can enhance the design of experiments and systems in collaborative VR embodiment.

Our findings reveal that research on collaborative embodiment in VR is highly varied, with methodologies reflecting a balance between laboratory-based experiments and field studies. Laboratory studies dominate due to their ability to control variables and measure precise outcomes, such as task completion time and synchronization metrics, whereas field studies provide valuable insights into real-world collaboration dynamics but are comparatively underexplored. The analysis further highlights the reliance on traditional metrics, such as task accuracy and user engagement, while advanced approaches like physiological synchrony and gaze tracking are emerging as promising indicators of collaboration quality. Notably, participant diversity remains a limitation, with insufficient representation of expert users and marginalized groups, raising concerns about the generalizability of findings. These gaps, along with the challenges in standardizing task design and metrics, emphasize the need for interdisciplinary and adaptive methodologies. This survey serves as a foundation for advancing collaborative embodiment in VR by identifying methodological strengths, addressing current limitations, and proposing pathways for inclusive and scalable future research.

\section{BACKGROUND}

The concept of embodiment in virtual reality (VR) has evolved from early theories of virtual presence, which emphasized the user’s sense of being immersed in a computer-generated environment. Initial studies on presence, such as those by Slater and Wilbur (1997), introduced the idea that immersion relies on the multimodal sensory input provided by VR systems to create a convincing sense of being ``inside'' the virtual world~\cite{slater1997framework}. Over time, embodiment research expanded beyond immersion to incorporate more specific aspects like self-location, agency, and body ownership—all of which are essential to understanding how users relate to their virtual bodies (Kilteni et al., 2012)~\cite{kilteni2012sense}. These components were originally developed to improve the user’s interaction with and control over the virtual environment in single-user VR setups, enhancing the sense of personal presence and natural engagement within the space.

In the field of Human-Computer Interaction (HCI), research on embodiment has been crucial for designing more intuitive and immersive VR systems. HCI frameworks like user-centered design have played a significant role in shaping how embodiment is understood, especially through the study of interaction design and how users perceive their avatars in VR. Studies have shown that the representation of the body in VR, including first-person perspectives and avatar synchronization, significantly impacts users' sense of control and immersion~\cite{gonzalez2019avatar}. These design advancements have been crucial for improving how users experience and interact with virtual environments, making them feel more present and engaged, even in multi-user or collaborative contexts.

\subsection{Design Approaches for Embodiment in VR}

\subsubsection{Self-Representation and Avatar Embodiment}
Genay~\cite{genay2021being} suggested that, self-representation in virtual environments is approached through various design methodologies that focus on the degree of similarity between the user's physical body and their virtual avatar. A key element in avatar design is the ability to customize the avatar to reflect the user's personal characteristics, such as body shape, clothing, and facial features, which increases identification and comfort~\cite{javornik2017magicface}. Research shows that users tend to feel more connected to avatars that resemble their real-world selves, which can enhance their overall experience in virtual environments~\cite{de2011embodiment}. However, more abstract or creative representations allow users to explore different identities, providing flexibility in self-representation~\cite{van2017presence}.
When discussing avatar embodiment, techniques such as first-person perspectives and visual feedback through mirrors are critical in fostering a sense of ownership over the virtual body~\cite{rosa2019embodying}. First-person perspectives allow users to see the virtual world through the eyes of their avatar, strengthening the feeling of being ``inside'' the virtual body~\cite{lenggenhager2007video}. Mirror reflections further support embodiment by allowing users to observe and control their virtual body, reinforcing the connection between the user and the avatar~\cite{harrison2011omnitouch}. Customization also plays an essential role, as giving users the ability to adjust their avatar's appearance to match their preferences enhances their sense of control and ownership over their virtual self~\cite{parfit1971personal}. These design elements collectively contribute to a more immersive and embodied virtual experience~\cite{stone2013appearance}.

\subsubsection{Shared Control and Multi-User Interactions}
In Shared Control, research has explored methods where multiple users share control over a single avatar, using techniques like the weighted-average co-embodiment method, which distributes control percentages between participants~\cite{fribourg2020virtual,hagiwara2019shared,kodama2022enhancing}. Studies show that increasing control weight improves users' sense of agency, enhancing task coordination~\cite{kodama2022enhancing}. This method has been effective in contexts like VR-based rehabilitation, where even users with less control feel engaged~\cite{juan2023immersive,gonzalez2017immersive}.

Further work in multi-operator single robot (MOSR) systems examined how multiple users control a robot, later evolving to include shared control in VR environments with a focus on first-person perspectives and multisensory feedback, affecting psychological aspects like intention alignment~\cite{inami2022cyborgs,jeunet2018you,kennedy1993simulator}. The phenomenon of ``enfacement'' where users perceive a merged identity, highlights the immersive potential of shared control~\cite{schubert2001experience}.

In learning and work settings, shared control has been shown to improve motor skill learning and task efficiency~\cite{moll2009communicative,mekbib2021novel}. Systems like Legion allow remote users to control interfaces collaboratively, focusing on utility and performance in highly controlled environments~\cite{mueller2020next}.

The effectiveness of multi-user collaboration in virtual environments is strongly influenced by mechanisms for co-presence, communication, and task synchronization. Nonverbal communication (NVC) is central to enhancing collaboration in multi-user virtual environments (MUVEs), as it facilitates more intuitive user interactions~\cite{zhou2024detecting}. Specifically, gaze behavior and body language have been shown to significantly impact the quality of collaboration, with effective gaze coordination enhancing both task synchronization and social engagement~\cite{schneider2021gesture}. The use of expressive avatars that convey facial expressions further contributes to the sense of social presence, which is crucial for successful collaboration~\cite{tarnec2023effect}. Moreover, spontaneous nonverbal coordination among users, facilitated by embodied avatars, has demonstrated potential in improving real-time task execution in collaborative virtual settings~\cite{parikh2014role}. These findings emphasize the importance of integrating nonverbal cues to support real-time collaboration in VR.

\subsubsection{Haptic Feedback and Sensory Integration}

Since the early days of VR, haptic feedback has been fundamental in enabling diverse touch-based interactions, with vibrotactile and force feedback being the most common types used in VR applications~\cite{srinivasan1997haptics,wee2021haptic,yang2021recent}. The integration of haptic feedback in shared virtual spaces has consistently enhanced user experiences and increased perceived social presence~\cite{jung2021use,oh2018systematic,zhang2023remotetouch}. In co-embodiment scenarios, studies by Hapuarachchi and Kitazaki explored the use of visual and passive haptic feedback to manipulate the sense of agency and maintain consistent posture~\cite{parikh2014role,hapuarachchi2022knowing}.
This body of work underscores the importance of feedback mechanisms in facilitating effective collaboration and shared control in multi-user virtual environments~\cite{fribourg2020virtual,kodama2022enhancing}.

\subsubsection{Immersive Interaction Design}

Natural User Interfaces (NUIs) in embodiment are critical in enabling intuitive user interactions within virtual environments, significantly enhancing user control and agency. The review by Bachmann et al. discusses how hand tracking, gesture control, and eye tracking, particularly through devices like the Leap Motion Controller, provide more natural and immersive user interactions in VR~\cite{bachmann2018review}. These interactions help users feel more embodied, making external devices feel like natural extensions of their bodies. Additionally, the survey by Zender and Wehrmann examines Environmental Interaction, highlighting how virtual environments are designed to naturally respond to user actions, such as interacting with objects, changing perspectives, and navigating virtual spaces~\cite{yeo2024entering}. These environmental interactions are key to creating an immersive and embodied VR experience, which is further supported by the use of NUIs that facilitate seamless and natural feedback loops~\cite{chen2024survey}.

\subsubsection{Social Presence and Co-Embodiment}

Enhancing social presence in multi-user virtual reality (VR) environments has been a significant focus in recent research. In their survey, Roth et al. analyze design approaches that enable users to feel physically co-located with others in the same virtual space, emphasizing realistic avatar representations and interaction cues~\cite{kyrlitsias2022social}. Schubert et al. explore how social presence affects cooperation in large-scale multi-user VR, highlighting methods to enhance social interaction through eye contact, facial expressions, and body language~\cite{wienrich2018social}. Slater et al. go beyond mere replication of physical behaviors by proposing techniques that augment social behaviors in VR, thereby enriching non-verbal communication and social interactions~\cite{roth2018beyond}. These studies collectively demonstrate the importance of detailed avatar representation and sophisticated interaction methods in promoting social presence and co-embodiment in VR environments.


\subsection{Methodological Concerns in HCI and Embodiment Research}

Methodological trade-offs play a crucial role in the field of Human-Computer Interaction (HCI) and embodiment research. While a comprehensive discussion of all methodological concerns is beyond the scope of this paper, we focus on five critical aspects that are particularly relevant to studying collaborative embodiment: sampling and representation of users, study environments, evaluation metrics, technological limitations, and ethical considerations.

\subsubsection{Participants}

The selection of participant populations is crucial for generalizing findings in collaborative embodiment research. Engaging participants with diverse demographics captures a wide range of user experiences during shared control tasks. For example, studies assessing embodiment in shared VR environments include users with varying levels of VR familiarity to assess the influence of shared environments on embodiment~\cite{zhou2024coplayingvr}. Balancing sample size and diversity ensures sufficient statistical power while maintaining diversity~\cite{latoschik2017effect}.
Incorporating control groups is essential to evaluate methodologies across different populations. Including control groups verifies whether results are consistent across demographics~\cite{cullen2021considerations}.

\subsubsection{Study Environment}

The Virtual Co-Embodiment study emphasizes that controlled lab environments offer greater precision in manipulating variables such as synchronization and shared control between users. However, this level of control may reduce ecological validity, as real-world factors like unpredictable physical environments are excluded, limiting the generalizability of the findings to actual collaborative VR applications~\cite{fribourg2020virtual}. Similarly, the ShareYourReality study highlights that while lab-based setups allow for detailed tracking of user interactions, more immersive environments are essential for studying how immersion affects collaboration dynamics in more realistic settings~\cite{venkatraj2024shareyourreality}.
However, lower levels of immersion, such as setups with limited sensory input, can diminish the sense of embodiment and reduce the effectiveness of collaborative efforts~\cite{kilteni2012sense,fribourg2020virtual}.

\subsubsection{Metrics}

In studies examining the metrics of embodiment in virtual reality (VR), both objective and subjective metrics are crucial for a comprehensive evaluation. Objective metrics often include task performance, such as accuracy, task completion time, and error rates, which provide quantifiable data on how well users can control and interact within the virtual environment. For example, in previous research, performance metrics were used to evaluate the success of user interaction within the virtual body~\cite{slater2010first,kilteni2012sense}. Subjective metrics, on the other hand, assess the user’s personal experience, including the sense of agency, body ownership, and overall user satisfaction, as highlighted in related studies~\cite{zhou2024coplayingvr}. Other works integrate both types of metrics to analyze how different task types and environments influence the user’s sense of embodiment, offering insights into how effective VR systems are in promoting realistic and engaging experiences~\cite{gonzalez2020self}. Additionally, research has further explored how technological embodiment can shape user experiences and engagement~\cite{garau2005responses}.

\subsubsection{Technological Limitations}

Hardware limitations, such as inadequate motion tracking, limited haptic feedback, and issues with field of view, restrict the user's ability to fully experience an embodied virtual presence~\cite{biocca1997cyborg}. Additionally, the software challenges include real-time synchronization difficulties in multi-user environments, which complicates collaborative embodiment experiences~\cite{roberts2004controlling}. These issues can reduce the sense of agency and body ownership, making it harder to create meaningful and engaging interactions~\cite{kilteni2012sense}.


\section{METHODOLOGY}

This review adheres to the four-phase guidelines outlined by the Preferred Reporting Items for Systematic Reviews and Meta-Analyses (PRISMA)~\cite{page2021prisma}. Initially, a search strategy was implemented to retrieve study records from publication databases (Section 3.1). Subsequently, titles and abstracts were screened to exclude studies that were not relevant to the review’s scope. In the third phase, the full-text articles were evaluated according to predefined eligibility criteria (Section 3.2). Ultimately, studies that met these criteria were selected for inclusion in the analysis.

\subsection{Search Strategy}

The initial search was performed based on the topic of this review to include the most relevant studies. Considering the two main parts of the topic—collaborative embodiment and virtual reality—the query statement was defined below:

\begin{verbatim}
(embodiment OR "body ownership" OR agency OR presence OR "co-embodiment")
AND
(collaborative OR collaboration OR "multi-user" OR shared OR "co-presence" OR "shared perspective")
AND
("virtual reality" OR VR OR "immersive environment*" OR "virtual environment*")
\end{verbatim}

The asterisk symbol (*) in the query represents arbitrary non-space characters. This query was crafted to capture studies discussing the general concept of embodiment as well as those focusing specifically on collaborative aspects within virtual reality environments. The search targeted titles and abstracts across four major databases: ACM Digital Library, IEEE Xplore, Springer Link, and Elsevier Scopus (including Science Direct). To ensure relevance, searches in the Springer and Elsevier databases were restricted to conference papers or journal articles in Human-Computer Interaction and Virtual Reality. The publication range was set from 2014 to 2024 to include recent studies. Notably, since Springer Link lacks an option to limit searches to titles or abstracts, we adopted a modified approach: first, performing a full-text search and then filtering results using a Structured Query Language (SQL) statement, identical to the queries applied in ACM, IEEE, and Elsevier databases.

\begin{figure}[t]
    \centering
    \includegraphics[width=0.6\textwidth]{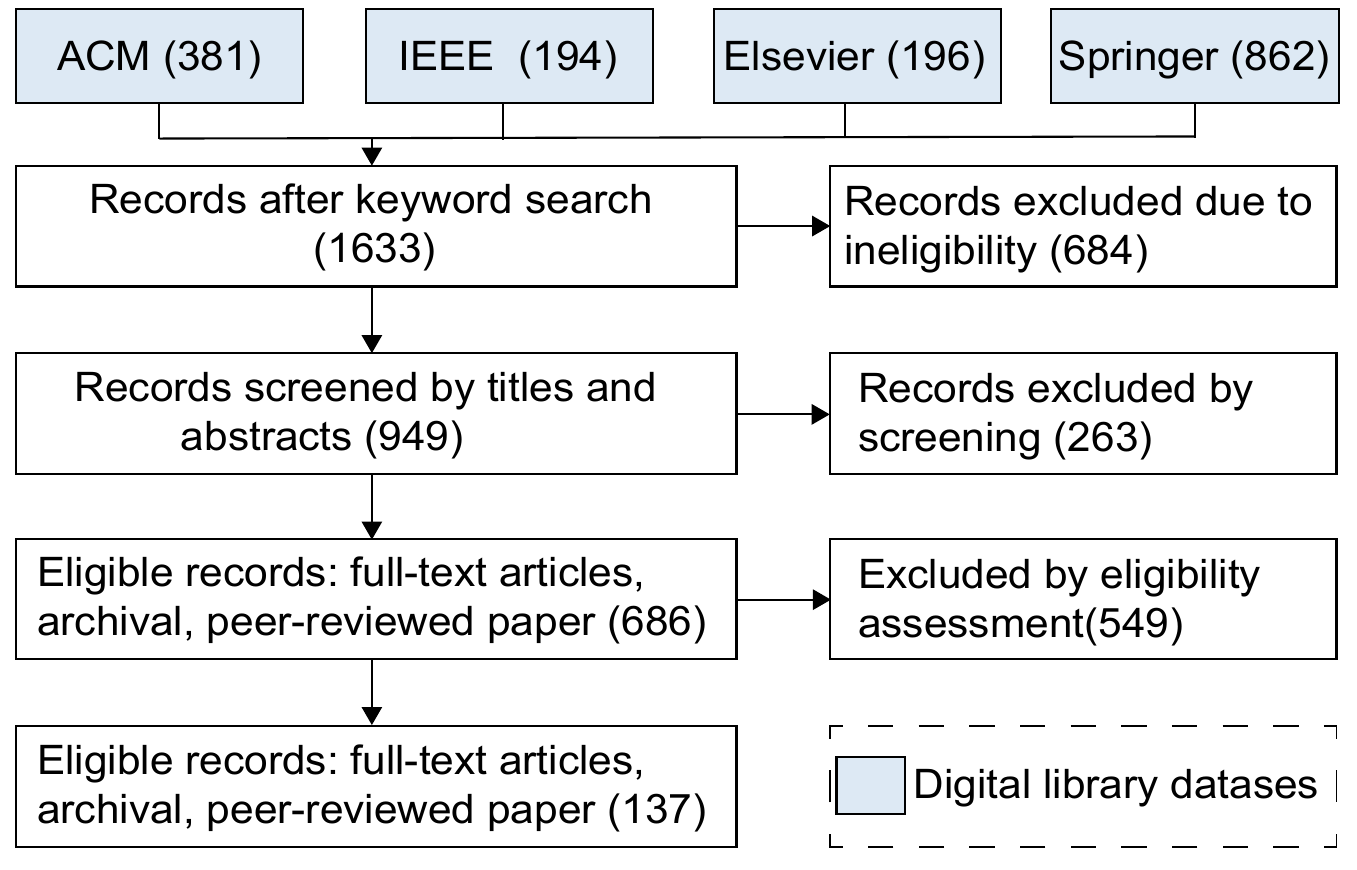} 
    \caption{PRISMA flow chart highlighting the stages of our systematic search}
    \label{Library_data}
\end{figure}

\subsection{Study Selection}

A total of 1,633 query results were retrieved from the databases (381 from ACM, 194 from IEEE, 862 from Springer, and 196 from Elsevier). After removing duplicates by title, 949 studies were initially included for screening and eligibility assessment. The selection criteria for these studies were as follows:

Criterion 1 — Publication category: The study must be published as a peer-reviewed technical paper. Publications such as progress reports, opinion pieces, or dissertations are excluded.

Criterion 2 — Involvement of virtual reality: The study must involve virtual reality. Studies using only other technologies, such as augmented reality (AR), extended reality (XR), or enhanced reality (ER), are excluded.

Criterion 3 — Collaborative embodiment: The study must involve collaborative embodiment. Studies focusing solely on single-user embodiment, such as experiences involving a single user's virtual body, are excluded.

Overall, 263 studies were excluded during title and abstract screening, and 549 studies were excluded following the full-text eligibility assessment. In total, 137 studies were included in this review for analysis.




\section{RESULTS}

In this section, we present the findings from our systematic review of collaborative embodiment in Virtual Reality (VR). Our analysis identified a total of 137 papers published between 2014 and 2024, encompassing 150 studies. The majority of papers contained a single study (127), while 7 papers included 2 studies, and 3 papers featured 3 or more studies, consistent with previously reported local HCI standards~\cite{caine2016local}. To provide deeper insights into their specific methodological and conceptual contributions, we evaluate and present each study individually, rather than grouping them by paper. The analysis focuses on the measures and metrics, study tasks, collaboration methods, and study parameters employed in the reviewed studies. By examining these elements, we aim to identify common practices, methodological strengths, and gaps in the current research, which will inform future studies and the development of more effective and inclusive collaborative VR systems.

\subsection{Measures and Metrics} 

\subsubsection{Subjective Measures} 

\begin{table*}[ht]
\centering
\caption{Overview of Questionnaires and Key Metrics Measured}
\label{tab:questionnaires}
\begin{tabular}{p{5cm} c p{6.3cm}}
\toprule
\textbf{Questionnaire} & \textbf{Percentage of Papers} & \textbf{Key Metrics Measured} \\ 
\midrule
Igroup Presence Questionnaire (IPQ) & 35\% & Spatial Presence, Involvement, Experienced Realism \\ 
Slater-Usoh-Steed Questionnaire (SUS) & 25\% & Sense of ``being there,'' immersion relative to reality \\ 
NASA Task Load Index (NASA-TLX) & 30\% & Mental Demand, Physical Demand, Temporal Demand, Performance, Effort, Frustration \\ 
System Usability Scale (SUS) & 20\% & Usability of VR interfaces \\ 
User Experience Questionnaire (UEQ) & 15\% & Attractiveness, Perspicuity, Efficiency, Dependability, Stimulation, Novelty \\ 
Networked Minds Social Presence Inventory & 18\% & Co-presence, Psychological Involvement, Behavioral Engagement \\ 
Social Presence Questionnaire (SPQ) & 22\% & Co-presence, Perceived Interaction Quality, Emotional Connectedness \\ 
Simulator Sickness Questionnaire (SSQ) & 40\% & Nausea, Oculomotor Symptoms, Disorientation \\ 
Presence Questionnaire (PQ) & 28\% & Sensory Fidelity, Control Factors, Distraction \\ 
Body Ownership Questionnaire (BOQ) & 12\% & Agency, Self-Location, Ownership \\ 
\bottomrule
\end{tabular}
\end{table*}

This section highlights key subjective measures used to evaluate collaborative embodiment in VR. Figure~\ref{Methods_Evaluation} summarizes how different collaboration methods are assessed across multiple criteria, reflecting diverse strategies for capturing user experience. The analysis of questionnaire usage in collaborative VR research reveals a strong emphasis on presence, social interaction, workload, usability, and embodiment, with distinct trends in how these aspects are measured. Across studies, presence-related metrics dominate, reflecting the central role of immersion in VR collaboration. Questionnaires such as the IPQ~\cite{schubert2001experience}, PQ~\cite{witmer1998measuring}, and SUS~\cite{usoh2000using} are widely used to assess users' sense of ``being there'' and how well the virtual environment replicates real-world interactions. Given that immersion directly influences engagement and task performance, these tools provide critical insights into system effectiveness, particularly when comparing VR-based collaboration with traditional methods.
Social presence is another key factor in evaluating collaborative VR experiences. Studies frequently employ the Networked Minds Social Presence Inventory~\cite{biocca2001networked} and SPQ~\cite{ratan2009self} to capture co-presence, emotional connectedness, and interaction quality. These metrics are especially relevant in multi-user VR environments, where the ability to perceive and respond to others in real time shapes overall collaboration quality. The reliance on these measures underscores the increasing recognition that effective VR collaboration extends beyond technical system design to encompass interpersonal dynamics.
Cognitive workload and usability assessments are also prevalent, often using NASA-TLX~\cite{hart2006nasa} and SUS~\cite{brooke1996sus} to balance system efficiency with user experience. NASA-TLX provides a multi-dimensional perspective on mental and physical effort, making it well-suited for analyzing complex collaborative tasks, while SUS remains a staple in evaluating interface usability. The integration of these tools suggests that researchers are not only concerned with how immersive VR is but also with whether it remains accessible, efficient, and easy to use, ensuring that collaboration is both effective and sustainable over extended periods.
While presence, social interaction, and workload represent primary evaluation categories, the measurement of embodiment is emerging as a distinct area of interest. The BOQ~\cite{botvinick1998rubber}, though less frequently employed, plays a crucial role in studies exploring body ownership and agency in shared VR experiences. Its application highlights a growing shift toward understanding how the perception of self-representation affects interaction, task execution, and collaboration quality.
Finally, simulator sickness remains a persistent concern in VR research, as evidenced by the widespread use of the SSQ~\cite{kennedy1993simulator}. Ensuring comfort in immersive environments is particularly crucial for collaborative VR systems, where prolonged exposure can exacerbate symptoms and negatively impact engagement. The frequent integration of SSQ suggests that minimizing motion discomfort is an active research priority, particularly for high-intensity or long-duration tasks.
Taken together, the choice of questionnaires in VR collaboration research demonstrates a balanced approach, integrating subjective measures of immersion, social presence, cognitive effort, usability, embodiment, and physical comfort. However, variability in questionnaire selection across studies presents challenges for direct comparisons, indicating a need for more standardized evaluation frameworks. 

\subsubsection{Objective Measures} 

We began by examining the objective measures employed in the different studies to evaluate collaborative virtual reality (VR) systems. Typically, these measures were used to assess user performance, interaction quality, and coordination within multi-user VR environments. From our analysis of the 113 studies, we identified a range of objective metrics. Among these, \textit{quality of experience (QoE)}~\cite{magalie2018toward,subramanyam2022evaluating,gunkel2021vrcomm,jansen2024open,olin2020designing,mai2018evaluating,seele2017here,liu2019supporting,osborne2023being,kolesnichenko2019understanding,magalie2018toward,gunkel2021vrcomm,jansen2024open,olin2020designing,mai2018evaluating,seele2017here,liu2019supporting,osborne2023being,kolesnichenko2019understanding} emerged as the most prevalent measure, with 20 studies (15\%) evaluating system responsiveness, latency, and frame rates to assess the technical stability and user experience of VR systems. Following this, \textit{gaze and gesture analysis}~\cite{venkatraj2024shareyourreality,adkins2024hands,kimmel2023let,wolf2022socialslider,li2024social,le2024impact,podkosova2018co,dobre2022more,seele2017here,andrist2017looking,vspakov2019eye,fidalgo2023magic,zhao20223d,davalos20243d} was reported in 15 studies (11\%), primarily used to track non-verbal cues and evaluate user attention, coordination, and communication behaviors. Similarly, \textit{task accuracy}~\cite{venkatraj2024shareyourreality,freiwald2021effects,kimmel2023let,le2022improving,kocur2020effects,freiwald2020conveying,vspakov2019eye,kodama2023effects,kangas2022trade,zhang2023towards,pouliquen2016remote,takita2024motor,kodama2022enhancing} and \textit{communication frequency}~\cite{venkatraj2024shareyourreality,adkins2024hands,kimmel2023let,wolf2022socialslider,smith2018communication,liu2018no,hai2018increasing,seele2017here,ghamandi2024unlocking,bayro2022subjective} were each employed in 15 studies (11\%). \textit{Task accuracy}~\cite{kodama2023effects,kangas2022trade,zhang2023towards,pouliquen2016remote,kodama2023effects,takita2024motor,kodama2022enhancing} evaluated the precision of task outcomes, making it ideal for skill-based tasks, while communication frequency captured both verbal and non-verbal exchanges to measure collaboration dynamics and engagement.

\begin{center}
\includegraphics[width=0.9\linewidth, height=0.35\textheight, keepaspectratio]{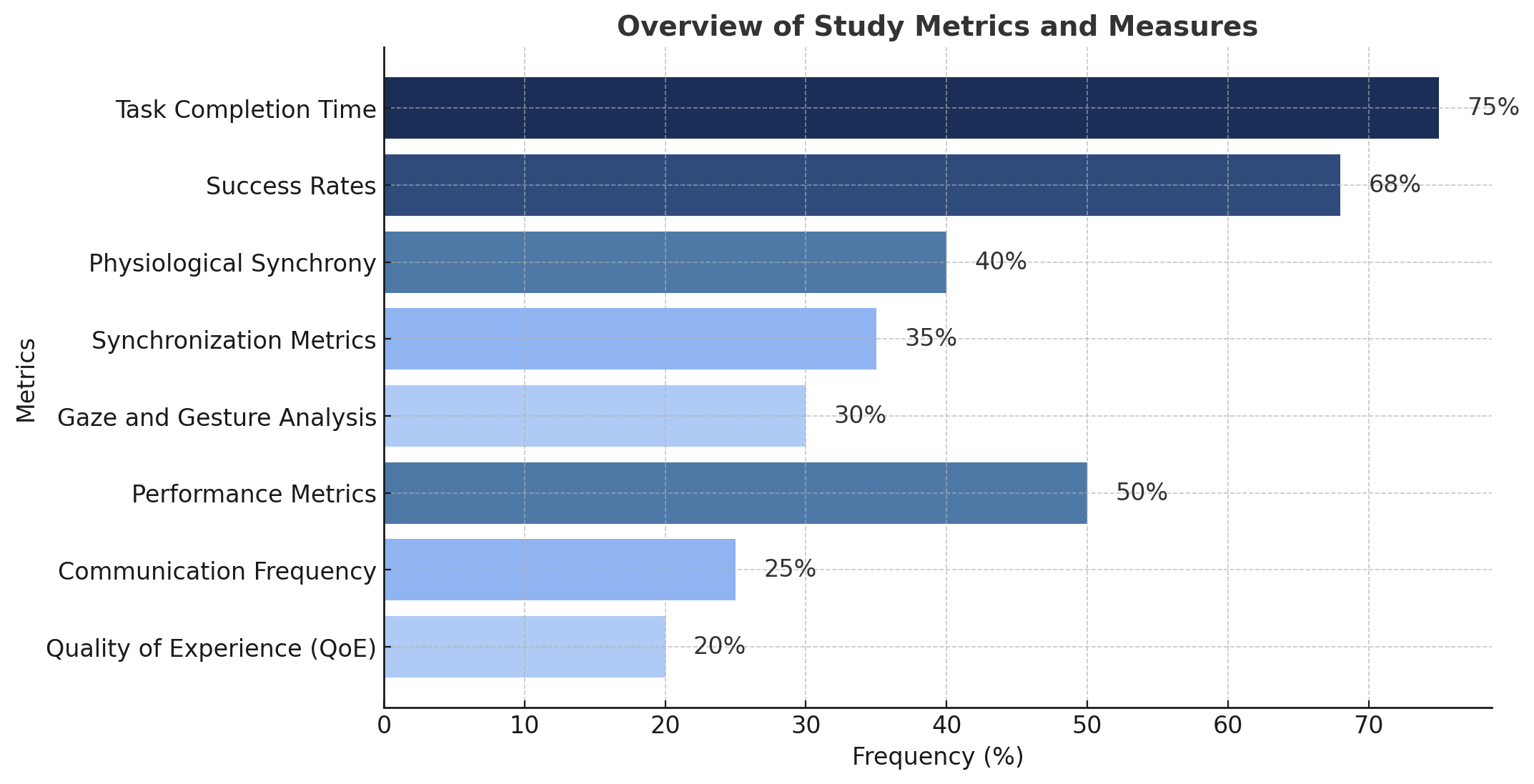}
\captionof{figure}{Objective Evaluation Matrix for Methods of Collaboration}
\label{Methods_Evaluation}
\Description{Objective Evaluation Matrix for Methods of Collaboration showing various metrics used to evaluate collaborative methods.}
\end{center}


Furthermore, \textit{physiological synchrony}~\cite{moharana2023physiological,salminen2019evoking,sasikumar2022pscvr,sun2019nonverbal,chuang2024leveraging,lee2020virtual,venkatraj2024shareyourreality,li2024social,fribourg2018studying,hagiwara2019shared,bozgeyikli2021give,takizawa2019exploring} was utilized in 13 studies (10\%), employing measures such as heart rate and electrodermal activity to assess shared engagement and emotional alignment between users. \textit{Task completion time}~\cite{adkins2024hands,freiwald2021effects,le2022improving,mai2018evaluating,freiwald2020conveying,lacoche2017collaborators,liu2019supporting,bayro2022subjective,dubosc2021impact,dominic2020remote,born2019co,kodama2023effects,jiang2023virtual}, a straightforward indicator of efficiency, was reported in 12 studies (9\%) to evaluate performance speed and collaboration efficiency. Likewise, \textit{synchronization}~\cite{venkatraj2024shareyourreality,wolf2022socialslider,zhou2019astaire,lu2024light,piitulainen2022vibing,bozgeyikli2021give,hagiwara2019shared,he2024mindmeld,xue2024collaborative,takizawa2019exploring,fribourg2018studying,kodama2023effects} between users was measured in 12 studies (9\%), particularly in tasks requiring precise timing or coordinated actions, such as shared control scenarios. Lastly, \textit{success or error rates}~\cite{adkins2024hands,freiwald2021effects,le2022improving,mai2018evaluating,freiwald2020conveying,lacoche2017collaborators,liu2019supporting,dubosc2021impact,dominic2020remote,born2019co,kodama2023effects} were included in 11 studies (8\%) to reflect accuracy and identify performance bottlenecks in collaborative tasks.

This analysis highlights the importance of selecting appropriate measures and metrics, as it directly impacts how future collaborative VR systems are evaluated and compared. Standardizing these choices will enable researchers to better benchmark collaboration performance and user experience in multi-user virtual environments.

\subsection{Study Tasks} 

\begin{figure}[t]
    \centering
    \includegraphics[width=0.9\textwidth]{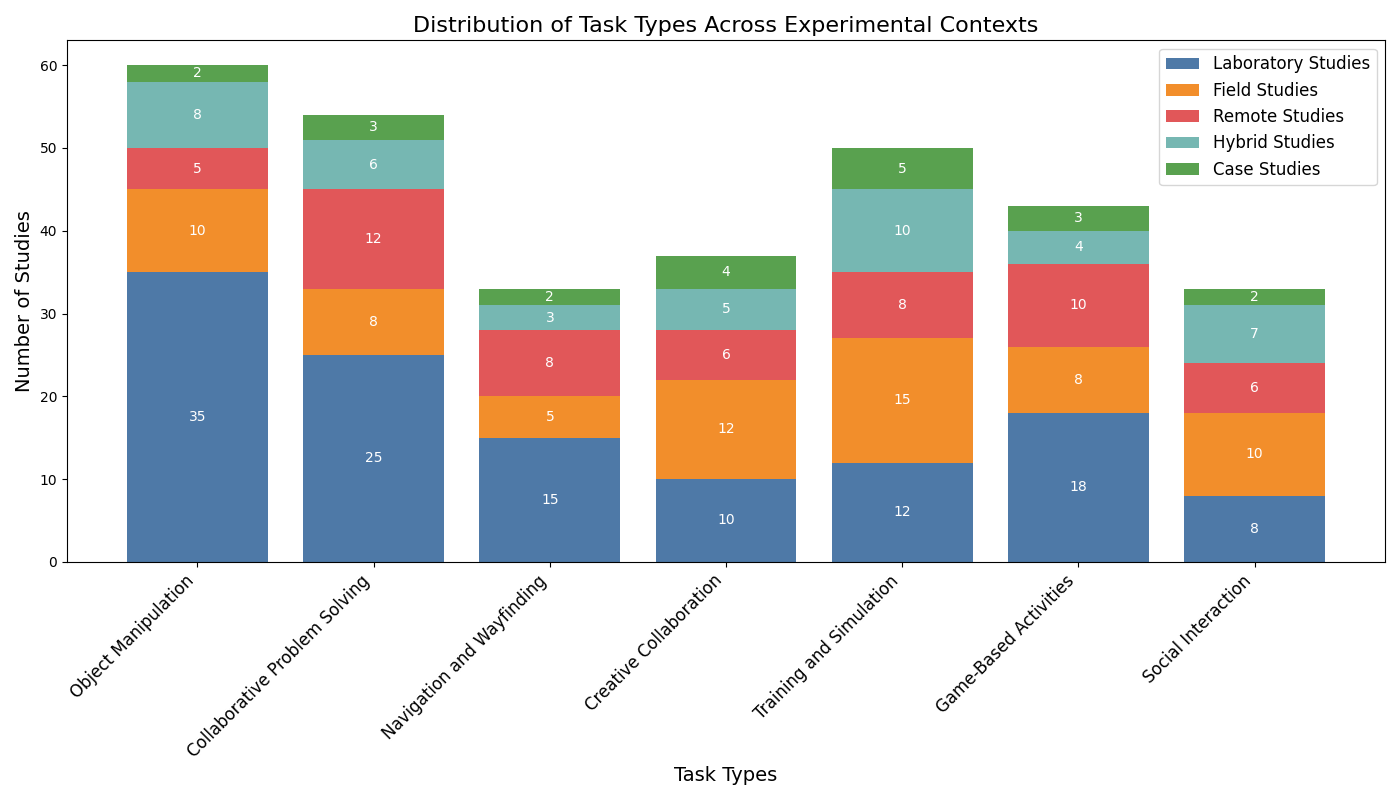} 
    \caption{Frequency Of Task Types In Collaborative VR Research}
    \label{Task_Types}
    \Description{A bar chart showing the frequency of different task types in collaborative VR research.}
\end{figure}

\begin{table*}[ht]
\centering
\caption{Overview of Task Categories, Specific Tasks, Measure Variables, and Study Distribution}
\label{tab:study_tasks}
\renewcommand{\arraystretch}{1.2} 
\begin{tabularx}{\textwidth}{>{\RaggedRight\arraybackslash}X 
                                    >{\RaggedRight\arraybackslash}X 
                                    >{\RaggedRight\arraybackslash}X 
                                    >{\RaggedRight\arraybackslash}X}
\toprule
\textbf{Task Category} & \textbf{Specific Tasks} & \textbf{Measure Variables} & \textbf{References} \\
\midrule
Object Selection and Activation & Selecting objects, pointing, clicking, highlighting, activating interface elements & Task completion time, accuracy, error rate & \cite{auda2021m, lakatos2014t, bozgeyikli2021give, freiwald2020conveying, kallioniemi2015collaborative, kuszter2014exploring, kuszter2015cooperation, gong2020interaction, conesa2023multi, huang2023spatially, petersen2023collaborative, almaree2024enhancing, wang2016exploring, kodama2023effects, jiang2023virtual, drogemuller2019remapping} \\
\midrule
Interface Interaction & Menu navigation, system configuration, file manipulation, tool usage & Usability, efficiency, user satisfaction, error rate & \cite{jiang2023virtual, ortiz2023workspace, ly2024collaxrsearch, ardal2019collaborative, nguyen2023collaborative, xia2018spacetime, zaman2015nroom, dorta2014hyve, nebeling2020xrdirector, mei2021cakevr, dominic2020remote, el2016cloud, fuchs2022collaborative, gong2020interaction, kumari2023exploring, pisoni2019interactive, takizawa2019exploring, sounti2022investigating, sasaki2022influence, maunsbach2023mediated, erfanian2017vibrotactile, hapuarachchi2022knowing, amastini2020collaborative} \\
\midrule
Object Manipulation and Assembly & Dragging and dropping, object assembly, fine motor tasks, force feedback interaction & Dexterity, task completion time, accuracy, force precision & \cite{auda2021m, lakatos2014t, bozgeyikli2021give, fidalgo2023magic, freiwald2020conveying, sugiura2018asymmetric, kallioniemi2015collaborative, kuszter2014exploring, kuszter2015cooperation, conesa2023multi, huang2023spatially, petersen2023collaborative, wang2016exploring, kodama2023effects, jiang2023virtual, drogemuller2019remapping, kumari2023exploring, pisoni2019interactive} \\
\midrule
Physical Activity-Based Tasks & Virtual sports, synchronized movement games, target shooting & Movement precision, engagement, task success rate & \cite{luo2021multi, zhou2019astaire, lu2024light, xue2024collaborative, he2024mindmeld, takita2024motor} \\
\midrule
Game-Based Tasks & Collaborative games, multiplayer strategy games, competitive gaming & Collaboration quality, engagement, task success rate & \cite{zhou2024coplayingvr, zhou2019astaire, dobre2022more, piitulainen2022vibing, boem2024takes, he2024mindmeld, xue2024collaborative, lu2024light} \\
\midrule
Collaborative Problem Solving & Resource management, joint construction, shared control tasks, decision-making activities & Time to solution, teamwork metrics, task success rate & \cite{zhou2019astaire, born2019co, dobre2022more, piitulainen2022vibing, he2024mindmeld, xue2024collaborative, li2024social, liu2018no, lu2024light, seele2017here, zhou2024coplayingvr} \\
\midrule
Creative and Expressive Tasks & Annotation, sketching, creating virtual objects & Creativity score, accuracy, task completion time & \cite{dorta2014hyve, lakatos2014t, nebeling2020xrdirector, ardal2019collaborative, lin2015space, mei2021cakevr, sounti2022investigating} \\
\midrule
Daily Virtual Activities & Sending emails, checking weather updates, scheduling, shopping & Usability, user satisfaction, completion time & \cite{ortiz2023workspace, gunkel2021vrcomm, osborne2023being, sharma2014multi, dominic2020remote, hube2021vr, el2016cloud, amastini2020collaborative, kumari2023exploring, pisoni2019interactive, zhou2024coplayingvr} \\
\midrule
Exploration and Navigation & Collaborative navigation, exploration games, coordinated navigation & Navigation efficiency, coordination, task success rate & \cite{kasahara2016parallel, born2019co, lee2020optimal, kuszter2014exploring, singh2020real} \\
\midrule
Simulation and Testing & Input device testing, simulation setup, system evaluation & Accuracy, device responsiveness, setup time & \cite{adkins2024hands, van2023haptic, venkatraj2024shareyourreality, zhou2024coplayingvr} \\
\bottomrule
\end{tabularx}
\end{table*}

We analyzed the types of tasks utilized in the studies to evaluate the performance of participants in collaborative virtual reality (VR) environments in Figure~\ref{Task_Types}. Among the 150 studies reviewed, most employed specific tasks to quantify user performance, interaction quality, and coordination, while a smaller subset relied on qualitative methods, such as interviews and surveys, without quantifiable tasks. Due to the diversity in task naming and definitions across the studies, we grouped tasks into broader categories based on their similarities. This process was conducted independently by two authors, achieving strong interrater reliability Cohen's Kappa ($\kappa$ = 0.91, $p < 0.01$).
The resulting categorization revealed ten primary task groups commonly used in the studies. Tasks such as selecting objects, pointing, clicking, and activating interface elements were grouped under ``Object Selection and Activation.'' Activities that involved interacting with or navigating virtual interfaces, including menu navigation, system configuration, and virtual file manipulation, were categorized as ``Interface Interaction.'' Tasks requiring manipulation or assembly of virtual objects, often involving fine motor skills, were grouped into ``Object Manipulation and Assembly.'' Tasks that emphasized physical movements, such as virtual sports or synchronized movement games, were classified as ``Physical Activity-Based Tasks.''

Several studies incorporated gameplay to evaluate coordination, decision-making, and strategy, which we categorized under ``Game-Based Tasks.'' Similarly, problem-solving tasks requiring teamwork, resource management, or shared control were grouped into ``Collaborative Problem Solving.'' Tasks encouraging creativity or expression, such as annotation, sketching, or creating virtual objects, were categorized as ``Creative and Expressive Tasks.'' Routine tasks adapted for VR environments, such as sending emails, checking weather updates, and scheduling, were grouped under ``Daily Virtual Activities.'' Tasks involving navigation or exploration in virtual spaces were categorized as ``Exploration and Navigation,'' while those focusing on testing or simulation scenarios, such as input device testing, were placed under ``Simulation and Testing.''

While most studies reported the use of at least one task to quantify user performance, approximately 23\% of the studies did not employ specific tasks, instead relying on qualitative insights from user feedback or observational data. This categorization highlights the diversity of tasks used in collaborative VR research, ranging from precise motor skill evaluations to routine and exploratory activities. A detailed overview of these tasks and their corresponding measure metrics is provided in Table~\ref{tab:study_tasks}, offering a foundation for standardizing task design in future research.

Understanding task design choices is crucial for future research on collaboration, as tasks shape the nature of interaction and the cognitive demands placed on users. Clear task frameworks will help researchers develop studies that more effectively capture the complexities of collaborative behavior in embodied virtual environments.

\subsection{Collaboration Methods}


In Table~\ref{tab:collaboration_methods}, we identified 23 collaboration methods across the reviewed studies, showcasing diverse approaches to teamwork in Virtual Reality (VR).
Shared Control allows multiple users to operate a virtual system collaboratively, but coordination complexity can increase with user numbers~\cite{fribourg2020virtual}. Multi-User Embodiment fosters a shared sense of presence through a single virtual avatar, though it may limit individual agency. Supernumerary Limbs~\cite{jiang2023virtual,drogemuller2019remapping,zhou2025juggling} enhance multitasking with additional virtual limbs but require significant cognitive adaptability.

Asymmetric Collaboration assigns distinct roles to participants, which may lead to imbalances if one role dominates~\cite{xia2018spacetime,clergeaud2017towards,sugiura2018asymmetric,kallioniemi2015collaborative,li2020omniglobevr,lakatos2014t,andrist2017looking,mei2021cakevr,dominic2020remote,fidalgo2023magic,zhao2023comparison,dunmoye2022exploratory,pisoni2019interactive,takizawa2019exploring,conesa2023multi}. Dynamic Role Switching offers flexibility in task assignments but risks disrupting task flow with frequent changes~\cite{venkatraj2024shareyourreality}. Distributed Collaboration supports remote teamwork but depends on stable network conditions, while Perspective Sharing enhances spatial understanding but can disorient users~\cite{venkatraj2024shareyourreality}.

Advanced methods, like Haptic Feedback Collaboration and Synchronized Multi-Modal Interaction, enrich interaction through sensory feedback but increase system complexity~\cite{adkins2024hands}. Tool-Driven Collaboration and Creative Collaboration enable co-creation but demand well-designed tools~\cite{ardal2019collaborative}. Gamified Collaboration~\cite{dobre2022more} promotes engagement through games but has limited use outside entertainment, while Interactive Decision-Making~\cite{nguyen2023collaborative} facilitates group consensus but can be time-intensive.

Joint Navigation~\cite{kallioniemi2015collaborative}, Resource Management~\cite{olaosebikan2022identifying}, and Collaborative Problem-Solving~\cite{sugiura2018asymmetric} emphasize teamwork in shared tasks but rely heavily on precise coordination. AI-Mediated Collaboration supports decision-making but may face trust challenges~\cite{ly2024collaxrsearch}. Real-World Context Integration bridges physical and virtual environments, though mismatches may disrupt immersion~\cite{wei2023bridging}.

Shared Physical Simulation~\cite{lu2024light} and Collaborative Learning~\cite{le2022improving} simulate realistic scenarios for skill development but depend on simulation quality. Interpersonal Communication Enhancement~\cite{kimmel2023let} strengthens social interaction but may be constrained by avatar fidelity, while Interactive Training~\cite{crowe2024modelling} targets skill-building in specific contexts but often requires significant customization.

Analyzing collaboration methods provides valuable insights into how different interaction designs support or hinder teamwork in VR. This knowledge is essential for guiding future system designs that aim to enhance coordination, role distribution, and shared control in collaborative virtual environments.

\begin{longtable}{p{0.165\textwidth} p{0.174\textwidth} p{0.174\textwidth} p{0.174\textwidth} p{0.21\textwidth}}
\caption{Collaboration Methods Overview}
\label{tab:collaboration_methods} \\
\toprule
\textbf{Collaboration Method} & \textbf{Primary Goal} & \textbf{Nature of Collaboration} & \textbf{Typical Setting} & \textbf{Task Examples} \\
\midrule
\endfirsthead

\toprule
\textbf{Collaboration Method} & \textbf{Primary Goal} & \textbf{Nature of Collaboration} & \textbf{Typical Setting} & \textbf{Task Examples} \\
\midrule
\endhead

\midrule
\multicolumn{5}{r}{\textit{Continued on next page}} \\
\midrule
\endfoot

\bottomrule
\endlastfoot

Shared Control & Task execution and system operation & Coordinated control of shared systems & Engineering, robotics, VR control systems & Joint manipulation of objects, shared avatar control \\
Multi-User Embodiment & Shared embodiment for task execution & Joint control of shared avatars & Avatar manipulation in VR & Synchronized movements \\
Supernumerary Limbs & Enhancing user capabilities with extra limbs & Collaboration with extended virtual limbs & Virtual multitasking environments & Multi-arm construction, virtual multitasking \\
Asymmetric Collaboration & Role-specific collaboration & Role-dependent tasks & Training scenarios, team-based tasks & Pilot and co-pilot collaboration \\
Distributed Collaboration & Remote collaboration & Collaboration across remote locations & Remote teamwork in VR & Remote co-design, telepresence \\
Avatar-Based Collaboration & Social interaction and task execution & Avatar-based interactions & Social VR platforms & Role-playing games, avatar communication \\
Haptic Feedback Collaboration & Improving tactile collaboration & Collaboration via shared tactile feedback & Haptics-enhanced VR systems & Force-feedback object manipulation \\
Perspective Sharing & Spatial alignment and shared perspective & Shared or swapped perspectives & Remote collaboration or co-design & Remote shared perspective tasks \\
Synchronized Multi-Modal Interaction & Combining sensory modalities for collaboration & Multi-sensory collaboration & Immersive VR experiences & Virtual music performance, storytelling \\
Tool-Driven Collaboration & Co-creation and problem-solving & Joint use of virtual tools & Design studios, virtual creative environments & Collaborative sketching or editing \\
Dynamic Role Switching & Dynamic adjustment of roles during tasks & Changing roles dynamically & Team-based dynamic tasks & Shifting leadership in tasks \\
Gamified Collaboration & Engagement and motivation & Task-based gaming scenarios & Gamified VR settings & Collaborative VR games \\
Real-World Context Integration & Integrating physical and virtual elements & Mixed-reality collaboration & Team-building and mixed-reality simulations & AR team-building exercises \\
AI-Mediated Collaboration & Facilitating collaboration with AI support & AI-enhanced interactions & AI-supported virtual collaboration & AI-mediated decision-making \\
Interactive Decision-Making & Consensus-building and decision-making & Collaborative decision-making & Strategy tasks, simulations & Collaborative design reviews \\
Joint Navigation & Co-navigation and exploration & Shared navigation tasks & Exploration of virtual spaces & Virtual museum navigation \\
Resource Management & Managing virtual resources & Collaborative resource handling & Resource-based simulations & Managing assets in virtual games \\
Creative Collaboration & Collaborative creativity and expression & Creative co-production & Virtual design and co-creation tasks & Creative VR art projects \\
Shared Physical Simulation & Task execution with virtual physics & Shared task execution with physics & Engineering simulations, virtual labs & Virtual assembly lines \\
Interpersonal Communication Enhancement & Enhancing communication quality & Improved social interactions & Social VR platforms, training tasks & Social gesture mirroring \\
Collaborative Learning & Learning and skill development & Joint learning activities & Educational and training settings & Solving virtual puzzles \\
Collaborative Decision-Making & Collaborative problem-solving & Team consensus and problem-solving & Team-based problem-solving & Strategy games \\
Interactive Training & Training for specific tasks & Practical skill development & Medical, military, or industrial training & VR firefighter training \\
\end{longtable}

\subsection{Study Parameters} 

\subsubsection{Participants} 

The reviewed 137 studies demonstrated notable variability in participant recruitment (Figure~\ref{Library_data}), with sample sizes ranging from 4 to 300 participants per study ($M = 22, Med = 18$). These sample sizes align with general practices in Human-Computer Interaction (HCI) research, though variations emerged based on study design and purpose. For instance, smaller exploratory studies involving qualitative assessments frequently recruited fewer participants (typically, $n < 10$), whereas studies focused on evaluating multi-user virtual reality (VR) systems and collaborative interactions often required larger sample sizes (n > 50).

Regarding participant demographics, the studies varied in reporting participants' VR experience. Novice VR users with limited or no prior experience appeared in 38\% of the studies, intermediate users with casual or short-term exposure were included in 42\%, and expert users with significant prior VR experience were featured in 20\% of the studies. Several studies included onboarding or familiarization sessions to mitigate performance differences due to varied expertise levels.

\begin{figure}[t]
    \centering
    \includegraphics[width=0.9\textwidth]{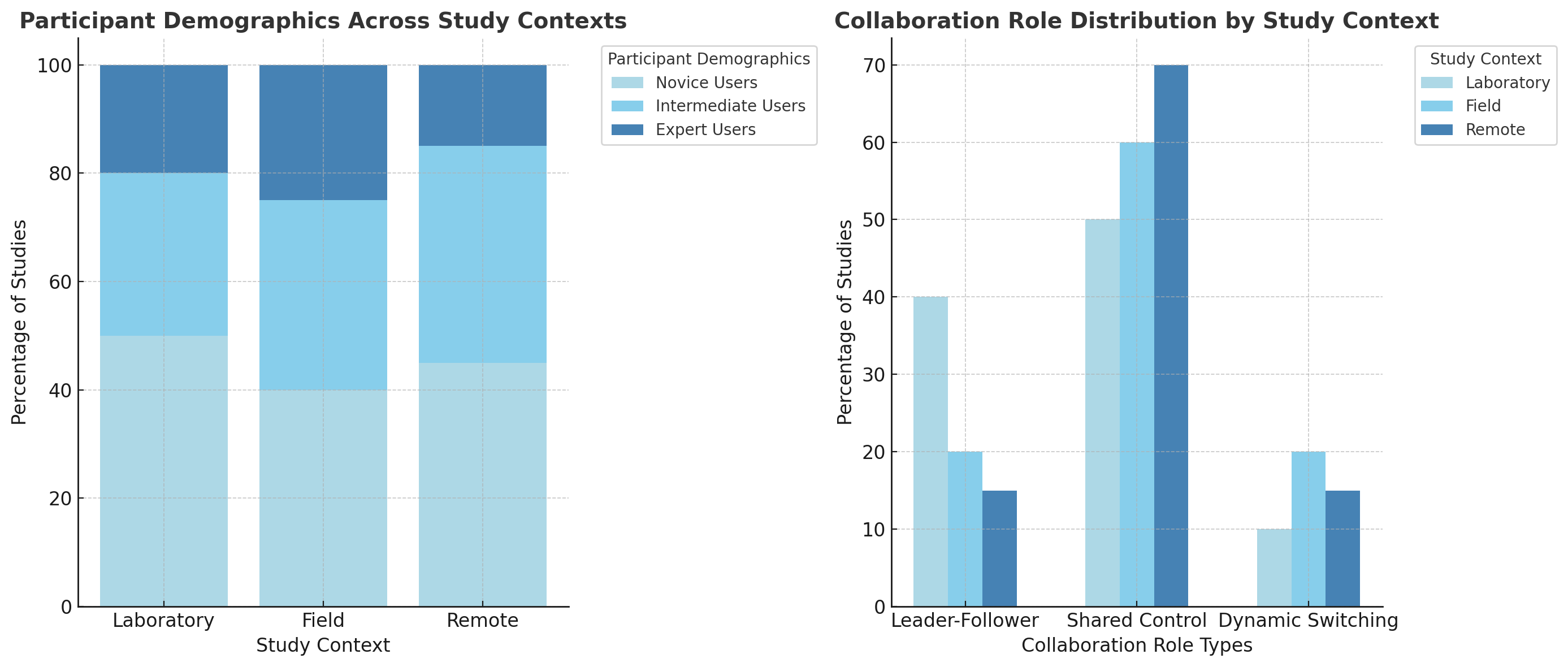} 
    \caption{Participant Demographics Across Study Contexts}
    \label{Library_data}
\end{figure}

The role distribution within collaborative tasks also differed across studies. Leader-follower roles were assigned in 30\% of the studies, where one participant assumed a guiding role (e.g., navigator or instructor). In contrast, shared control roles, where all participants had equal control over the task, were implemented in 62\% of the studies. Notably, 8\% of the studies reported dynamic role switching, where participants transitioned between roles based on task progression or emergent team strategies. This highlights the importance of role design in understanding collaborative dynamics and task performance.

Examining study parameters, including participant demographics and technological tools, helps future researchers understand the practical constraints and design decisions that influence collaborative outcomes. This perspective is critical for creating more inclusive, scalable, and realistic collaborative VR studies.

\subsubsection{Technological Tools and Systems} 

\noindent \textbf{\textit{Hardware Platforms:}} The studies reviewed utilized a range of VR hardware to support collaborative embodiment. Head-mounted displays (HMDs) were the most frequently employed devices (e.g.,~\cite{xia2018spacetime}) , appearing in 76\% of the reviewed studies (106 out of 137). In contrast, standalone devices like Meta Quest (e.g.,~\cite{zhou2024coplayingvr}) were favored in 28\% of studies (39 out of 137) for their portability and ease of deployment in real-world collaborative applications.

Haptic devices were integrated into 34\% of studies (47 in total) to provide tactile feedback and improve immersion (e.g.,~\cite{venkatraj2024shareyourreality}). Wearable solutions such as SenseGlove and Teslasuit were used for force and vibrotactile feedback in social interaction and remote collaboration experiments. However, due to cost and technical limitations, vibration-based actuators~\cite{erfanian2017vibrotactile} were more commonly employed to simulate force feedback in collaborative tasks.

Beyond standard VR peripherals, 19\% of studies (27 in total) incorporated non-traditional interfaces. These included robotic proxies for remote presence (e.g., ~\cite{sakashita2023vroxy}), multi-user redirected walking simulators (e.g., ~\cite{tomar2019conformal}), and custom tangible interfaces (e.g., ~\cite{dorta2014hyve}) designed to facilitate co-embodied interactions.

\noindent \textbf{\textit{Software Platforms:}}
VR software ecosystems varied across studies, ranging from commercial engines to custom-built frameworks. 59\% of studies (83 in total) leveraged real-time multi-user platforms such as Unity3D with Photon PUN or Unreal Engine for collaborative VR experiences(e.g., ~\cite{zhou2024coplayingvr}). These platforms offered scalability but required extensive customization for accessibility and heterogeneous device integration.

Specialized toolkits were utilized in 21\% of studies (29 in total) to support domain-specific applications. For example, Hyve-3D~\cite{dorta2014hyve} and Spacetime~\cite{xia2018spacetime} facilitated 3D sketching and fluid editing, while REVERIE integrated AI-driven avatars for adaptive interactions.

Web-based solutions appeared in 14\% of studies (20 in total), using WebRTC~\cite{jansen2024open} for browser-based social VR, enabling remote participation with reduced graphical fidelity. Systems such as VRComm~\cite{gunkel2021vrcomm} and Space Connection~\cite{lin2015space} balanced accessibility with performance trade-offs.

\subsubsection{Study Settings} 

The reviewed studies showed diverse methodologies and settings  (Figure~\ref{Library_data}). Laboratory-based studies dominated (67\%), leveraging controlled environments for tasks like comparing VR features or exploring embodiment effects. Field-based studies (17\%) occurred in naturalistic settings, including museums and collaborative workspaces, focusing on real-world applicability.

Incentives were reported in 34\% of studies, often involving monetary rewards or VR-related items, particularly in studies requiring extended or specialized participation. Methodologically, 39\% were controlled experiments comparing VR techniques, while 17\% were case studies describing unique VR applications.

Additionally, 10\% employed longitudinal designs to observe long-term trends, and 10\% focused on remote collaboration or VR meetings, reflecting a growing interest in distributed VR systems. This range highlights the field's focus on balancing experimental rigor with real-world relevance.

\section{DISCUSSION}

\subsection{Implications for Collaborative Embodiment} 

The study of collaborative embodiment goes beyond technical feasibility, revealing how shared virtual interactions reshape teamwork. This section synthesizes theoretical insights and practical challenges in multi-user VR, bridging empirical findings with real-world applications. While synchronization metrics and nonverbal cues inform teamwork dynamics, issues like cognitive overload, accessibility, and engagement highlight the need for adaptive design. By analyzing behavioral drivers of collaboration (e.g., physiological synchrony, role flexibility) alongside systemic barriers (e.g., hardware limits, cultural biases), we propose strategies to translate embodiment theories into scalable solutions. The following subsections explore these implications, advocating for a holistic approach to advancing collaborative VR systems.

\subsubsection{Insights into Collaborative Dynamics} 
We first highlight key insights that specifically reflect collaborative dynamics unique to collaborative embodiment, focusing explicitly on how multi-user interactions shape shared VR experiences. Our analysis highlights the critical role of synchronization, communication, and task-sharing mechanisms in designing multi-user VR systems. Metrics such as task completion time and success rates are widely used to assess collaboration, but deeper insights emerge from physiological synchrony and co-presence metrics, which reflect shared engagement and emotional alignment. For instance, studies show that higher synchronization in gaze or heart rate correlates with improved coordination and team cohesion~\cite{fasold2021gaze}.

However, current systems often overlook the potential of these metrics for real-time adaptability~\cite{novak2022harnessing}. Dynamic feedback mechanisms, such as adjusting task difficulty or communication tools based on physiological or behavioral data, remain underutilized. Similarly, while co-presence metrics capture the sense of "being together," their application seldom considers the impact of cultural or social diversity on collaborative dynamics~\cite{fairclough2017physiological}.

Existing research emphasizes the value of transparent communication and flexible role distribution in teamwork~\cite{hadziahmetovic2022role}. VR systems could incorporate these elements through features like real-time progress visualizations or shared task indicators, enhancing mutual awareness and coordination~\cite{rinnert2023can}.

Future research should investigate how collaborative dynamics evolve over time, particularly in complex tasks. Longitudinal studies could reveal how synchronization and communication adapt across repeated interactions, offering insights for scalable and inclusive VR systems that support diverse collaborative scenarios, including education, healthcare, and remote work.

\subsubsection{Practical Challenges and Solutions} 

Here we focus specifically on practical challenges unique to collaborative embodiment contexts, providing targeted solutions to improve multi-user collaboration in VR. Collaborative embodiment in VR faces practical challenges, such as managing cognitive load, ensuring accessibility, and sustaining engagement in long-term collaboration. High cognitive load, especially in tasks requiring real-time synchronization and decision-making, can overwhelm participants, particularly novice users~\cite{bueno2021effects}. Accessibility remains a concern, as many systems fail to accommodate users with diverse physical or cognitive abilities, limiting inclusivity~\cite{creed2024inclusive}. Maintaining engagement in prolonged tasks or longitudinal studies also poses significant difficulties, often leading to participant fatigue or dropout~\cite{budin2024participant}.

Adaptive interfaces and AI-driven task moderation offer promising solutions. Adaptive systems could dynamically adjust task complexity based on real-time metrics, such as physiological synchrony or error rates, ensuring an optimal balance of challenge and manageability~\cite{chiossi2023exploring}. Personalized onboarding and guidance tailored to individual expertise levels can help reduce cognitive load during initial interactions, AI-driven moderation can further enhance collaboration by analyzing interaction patterns and suggesting task or role adjustments to improve team efficiency~\cite{jony2024empowering}.

Engagement challenges could be mitigated through gamification and real-time feedback, which motivate participants by emphasizing progress and fostering a sense of achievement. Additionally, incorporating diverse accessibility features, such as voice commands and customizable controls, can make VR systems more inclusive~\cite{fernandez2024hands}. Future research should focus on refining these strategies to develop scalable and adaptable VR systems that cater to diverse user needs while ensuring sustained collaboration.

\subsection{Methodological Reflections on Collaborative Embodiment} 
This section specifically reflects on methodological aspects unique to collaborative embodiment research, examining how experimental design, contextual diversity, and measurement strategies collectively influence our understanding of multi-user VR scenarios. By comparing controlled laboratory studies with field experiments, we highlight the tensions between internal validity and ecological relevance. Additionally, we discuss evolving metrics and participant dynamics that challenge traditional paradigms, emphasizing the need for adaptive methodological frameworks that balance precision with inclusivity. The following subsections offer detailed insights to guide future innovations in collaborative embodiment methodologies.

\subsubsection{Study Design and Metrics} 
A major methodological gap in collaborative embodiment research is the scarcity of longitudinal studies~\cite{schubert2001experience}. Capturing how teams refine synchronization, adapt task strategies, and maintain engagement over time is essential for understanding collaborative dynamics in VR. Such insights are critical for designing systems that support sustainable and scalable teamwork~\cite{kothgassner2020does}. Longitudinal research would enable the study of learning effects, evolving role distributions, and the development of coordination patterns, providing a more realistic perspective on collaborative performance~\cite{parsons2015virtual}.
Another priority is the standardization of evaluation measures and metrics~\cite{dich2018using}. Our review shows significant variability in how collaboration quality and embodiment are assessed, making cross-study comparison difficult. Future studies should adopt consistent performance and subjective metrics tailored to collaborative contexts, such as measuring synchronization, shared workload, and social presence in addition to traditional task accuracy.
Finally, task design itself demands methodological attention. Carefully crafted collaborative tasks that reflect real-world complexity are necessary to elicit meaningful interaction patterns. Over-reliance on simplified tasks may mask the challenges of coordination and agency sharing, limiting the ecological validity of findings.

\subsubsection{Diversity in Experimental Contexts} 

The dominance of laboratory-based studies (67\%) in collaborative embodiment research ensures controlled conditions for isolating variables and measuring metrics like task efficiency and coordination~\cite{ghamandi2024unlocking} in current studies. However, this emphasis often limits the applicability of findings to real-world contexts, such as collaborative workplaces or public VR systems~\cite{kothgassner2020does}. In contrast, field studies (17\%) offer valuable insights into authentic interactions within dynamic environments but face challenges like smaller sample sizes and reduced control~\cite{hube2021vr}. Remote and hybrid collaborations, though promising, remain underexplored in terms of scalability and implementation.

Incentives, used in 34\% of studies, played a significant role in sustaining engagement, particularly in field and longitudinal studies. While monetary compensation or symbolic rewards (e.g., VR-related items) are effective, their impact on motivation and generalizability requires further examination~\cite{andrist2017looking}.

Future work should balance controlled and field studies, leveraging hybrid designs to combine rigor with ecological relevance. Addressing gaps in incentive strategies will support more inclusive and scalable methodologies, broadening the practical applicability of collaborative embodiment research.

\subsection{Technological and Methodological Gaps Specific to Collaborative Embodiment} 
Collaborative embodiment research has advanced through technological innovations and interdisciplinary insights, yet challenges remain in balancing inclusivity, scalability, and methodological rigor. While haptic feedback, adaptive avatars, and cross-platform integration showcase VR’s potential, issues like user diversity, inconsistent task design, and technical bottlenecks limit broader applicability. By contrasting achievements—such as improved co-presence metrics and AI-mediated collaboration—with gaps in participant representation and methodological standardization, we highlight tensions between technological ambition and practical implementation. The following subsections examine these strengths and limitations, offering strategies to align innovation with inclusivity and enhance research accessibility.

\subsubsection{Technological Advances and Applications}

The integration of advanced VR technologies, such as full-body tracking, haptic feedback, and adaptive avatars, has significantly enhanced collaborative embodiment research. These innovations enable researchers to simulate nuanced interactions, such as role transitions and synchronized team activities, and facilitate applications across education, healthcare, and remote collaboration. However, persistent technical bottlenecks undermine their full potential. Despite these challenges, recent studies such as the Whisper system demonstrate effective multisensory solutions that significantly enhance communication privacy and realism, thereby positively influencing users' social presence and collaborative interactions~\cite{wang2025vr} Motion capture systems (e.g., Vive Trackers, OptiTrack) struggle with tracking subtle movements like finger gestures and tremors, leading to inaccuracies in avatar control and interaction precision [Hands or Controllers~\cite{fribourg2020virtual}, Social Simon Effect in VR~\cite{yun2023animation}. Network latency in multi-user systems further disrupts real-time synchronization, degrading co-presence and skewing performance metrics such as task completion time~\cite{zhou2024coplayingvr}, Collaborative Motion Modes in Serious Game~\cite{he2024mindmeld}. Recent advancements in hardware acceleration and machine learning optimizations offer promising avenues to significantly reduce latency and enhance computational performance, thereby supporting more responsive and immersive collaborative VR experiences~\cite{fan2021high,ferianc2021improving}. Similarly, haptic feedback devices (e.g., Teslasuit, SenseGlove) face constraints in resolution and force range, failing to simulate nuanced tactile interactions like texture differentiation, which affects user agency and behavior in collaborative tasks~\cite{venkatraj2024shareyourreality}.

Beyond hardware constraints, cross-platform integration presents significant challenges. Heterogeneous VR systems introduce fidelity mismatches—combining untethered mobility with high-fidelity systems creates disparities in tracking accuracy and rendering, leading to spatial disorientation in shared environments~\cite{smith2019virtual}. Mixed input modalities, such as HTC Vive controllers versus touchscreen tablets, exacerbate interaction inequalities, as lower-fidelity interfaces restrict participation and control precision~\cite{luong2023controllers}. Without standardized protocols, these inconsistencies hinder the scalability and generalizability of findings.

Another critical gap is the lack of adaptive VR systems tailored to diverse user needs. Few studies have explored long-term adaptability, particularly in evolving user contexts such as accessibility for individuals with motor or cognitive impairments. This limitation raises concerns about VR’s scalability for large-scale applications in remote collaboration and distributed learning scenarios~\cite{biswas2021adaptive}. Future research must prioritize sensor accuracy, latency reduction, and adaptive haptics, alongside open-source frameworks (e.g., VR2Gather~\cite{jansen2024open}), to bridge these technological gaps. By improving cross-platform compatibility and ensuring inclusivity, VR can better align with real-world collaboration dynamics and expand its role in embodiment research.

\subsubsection{Limitations in Participant Representation} 

Participant representation and role dynamics are particularly crucial in collaborative embodiment. This subsection explicitly discusses how these factors uniquely influence shared VR experiences and outcomes. Our survey identifies notable gaps in participant representation in collaborative embodiment studies. Expert users, such as researchers or frequent VR users, are underrepresented, limiting the field’s ability to incorporate advanced feedback on usability and interaction techniques~\cite{peck2021divrsify}. Additionally, cultural diversity is rarely documented, raising concerns about the generalizability of findings to global user populations. This lack of inclusivity restricts the development of systems that address diverse user behaviors and expectations~\cite{do2024cultural}.

Insufficient onboarding strategies for novice users present another limitation. While many studies involve participants with limited VR experience, few describe detailed onboarding protocols to ease learning curves and reduce cognitive load. This omission risks introducing bias in task performance data, especially in complex collaborative tasks. Comprehensive onboarding could enhance engagement and standardize participant performance, yet its absence is a missed opportunity in much of the reviewed research~\cite{cesario2019boarding}.

Moreover, multi-role dynamics, such as leader-follower or shared control roles, remain underexplored. Although some studies acknowledge role assignments, transitions between roles and their influence on collaboration and satisfaction are rarely analyzed. Prior research emphasizes the importance of adaptable roles in improving coordination and task outcomes, highlighting a critical gap in current methodologies~\cite{lamon2023unified}.

To improve future research, participant recruitment should prioritize both expert and novice users from diverse cultural backgrounds to ensure inclusivity and ecological validity. Standardized onboarding protocols must also be developed to support novice users. Finally, more rigorous analysis of role dynamics is essential to understand their impact on collaboration and task performance, paving the way for more equitable and generalizable findings in collaborative VR systems.

\subsubsection{Inconsistencies in Methodological Standards} 

A lack of standardization in task design, metrics, and collaborative role distribution in studies on collaborative embodiment in VR. Task designs range from abstract experimental setups to domain-specific activities, creating inconsistencies that hinder comparability. Metrics such as task completion time, success rates, and physiological synchrony, while valuable for evaluating collaboration, are inconsistently applied, complicating cross-study comparisons. Similarly, collaborative role distributions—spanning fixed leader-follower roles, shared control, and dynamic role-switching—often lack clear justification or analysis of their impact on performance.

These inconsistencies pose significant challenges for replication. Variability in experimental protocols, measurement techniques, and analysis frameworks makes reproducing results difficult, undermining reliability~\cite{usui2021meta}. For instance, critical details such as task parameters, onboarding processes, or VR system calibration are frequently underreported, limiting transparency and comparability~\cite{chauvergne2023user}. This methodological diversity impedes the accumulation of cohesive knowledge and the application of findings to real-world scenarios.

Fields such as accessibility research and HCI have demonstrated the benefits of standardized methodologies, offering a potential roadmap for VR research~\cite{mack2021we}. For example, developing a shared taxonomy for collaborative roles and tasks would enhance methodological clarity and facilitate replication~\cite{ghamandi2023and}. Similarly, establishing benchmark metrics and transparent reporting practices could unify future studies and create a foundation for broader insights.

Addressing these gaps requires prioritizing standardization and transparency. By adopting consistent task designs, metrics, and reporting frameworks, researchers can improve the reproducibility of findings and enable meaningful comparisons. These efforts are essential for advancing the study of collaborative embodiment in VR and translating research insights into practical applications.

\subsection{Future Directions for Collaborative Embodiment Research} 
The future of collaborative embodiment research calls for methodological refinement and societal relevance. While progress has been made, key challenges remain in long-term adaptation, interdisciplinary integration, and equitable design. This section explores expanding experimental paradigms, such as longitudinal studies and inclusive system design, while leveraging neuroscience, AI, and human-robot interaction to develop novel metrics and adaptive frameworks. Ensuring scalable architectures and diverse user cohorts will better align VR systems with global collaboration needs. The following subsections propose strategies to bridge these gaps, balancing innovation with ethical responsibility and accessibility.
\subsubsection{Expanding Experimental Scope} 
Experimental research on collaborative embodiment in VR has largely focused on short-term, task-specific studies, limiting insights into long-term learning, evolving team dynamics, and adaptation to shared control~\cite{sayadi2024feeling}. Longitudinal studies are crucial to understanding how users refine synchronization, develop collaborative strategies, and maintain engagement over time. Such research could also reveal the sustained effects of VR on cognitive and emotional states~\cite{khojasteh2021working}.

Expanding the experimental scope further requires interdisciplinary approaches, integrating insights from fields like neuroscience, psychology, and education. These disciplines could introduce advanced metrics, such as brain-computer interfaces for analyzing shared cognitive states or task designs aimed at fostering collaborative learning~\cite{pradeep2024neuroeducation,papanastasiou2020brain}. Additionally, collaborations with robotics and AI research could lead to adaptive systems that respond to team dynamics in real-time, enhancing both user experience and task outcomes~\cite{natarajan2024mixed}.

By incorporating methodologies from diverse domains, future research can explore collaborative embodiment in broader contexts such as healthcare, remote work, and large-scale virtual communities. This interdisciplinary approach is essential to ensure that VR systems are both effective and scalable, advancing theoretical understanding and practical applications across various real-world scenarios.

\subsubsection{Enhancing Inclusivity and Scalability} 

Current studies on collaborative embodiment in VR often lack participant diversity, with limited representation across cultural, professional, and experiential backgrounds~\cite{peck2021divrsify}. Broadening participant demographics, such as including individuals from underrepresented groups or experts with extensive VR experience, would help generalize findings and better address diverse user needs~\cite{do2024cultural}. Additionally, scaling VR systems to accommodate larger, distributed teams remains underexplored, despite its critical importance for applications in education, healthcare, and industrial collaboration.

Integrating AI-driven evaluation and adaptive systems presents a promising avenue to tackle these challenges. AI algorithms can dynamically adjust task roles, redistribute workloads, and monitor team dynamics in real time, ensuring equitable participation and improving collaboration efficiency~\cite{jo2024cognitive}. For example, adaptive role assignment could help novice users transition into leadership roles, while maintaining overall task performance~\cite{han2008multilevel}.

Moreover, scalable VR platforms could benefit from distributed architectures, enabling seamless interactions across geographically dispersed teams. These platforms can leverage machine learning to analyze large-scale behavioral data, identifying patterns to optimize collaboration strategies and enhance user experience~\cite{garcia2019collaborative}. By focusing on inclusivity and scalability, future research can unlock VR's potential to support diverse, large-scale collaboration in both experimental and real-world scenarios.

\section{LIMITATIONS}

Our survey has several limitations. First, it is strictly limited to the results of our search within the domain of collaborative embodiment in virtual reality. Although we aimed to conduct an extensive review, some relevant papers might have been excluded due to the use of alternative keywords or their publication in interdisciplinary venues outside computer science (e.g., neuroscience or psychology). Furthermore, we focused on peer-reviewed academic literature, which may have resulted in the exclusion of valuable insights from industry reports or non-peer-reviewed studies. While we systematically classified existing methodological approaches, this classification might oversimplify hybrid or unconventional frameworks, potentially misrepresenting studies that do not fit neatly into predefined categories. Moreover, the rapid pace of technological development in virtual reality may limit the long-term applicability of some findings discussed in this survey. Finally, our survey focused on collaborative contexts, leaving out individual embodiment techniques or their integration with collaborative mechanisms, which could provide a more comprehensive understanding of embodiment in VR. These limitations highlight opportunities for future research to expand and refine the insights presented in this survey.


\section{CONCLUSION}

This paper presents a comprehensive survey of methodological approaches to collaborative embodiment in Virtual Reality (VR). Our review analyzed a diverse set of studies focusing on how collaborative embodiment is operationalized and measured within multi-user VR environments, covering metrics, participant interaction methods, and technological tools designed to enable collaborative experiences. Our findings reveal key challenges in standardizing methods for evaluating embodiment, including the lack of consistent metrics, the limited exploration of diverse collaboration contexts, and the underrepresentation of longitudinal studies in the field. We systematically categorized existing methodological approaches and provided a structured overview to guide researchers in designing and evaluating future collaborative embodiment studies.

Our survey identifies three critical research gaps: the need for a unified set of metrics to evaluate embodiment across different collaboration contexts, the necessity to explore cross-disciplinary methodologies to enhance the richness of VR studies, and the importance of investigating long-term impacts of collaborative embodiment in VR on user performance and experience. While current methods have provided valuable insights, further development is needed to enable accessible, scalable, and inclusive VR systems for diverse user groups.

This work offers a foundation for future researchers and designers seeking to enhance collaborative embodiment in VR by providing a structured overview of existing practices, challenges, and opportunities. It also highlights the need for interdisciplinary collaborations and iterative methodologies to deepen the understanding of how collaborative embodiment directly impacts teamwork, interaction, and shared control in virtual environments. By addressing the identified gaps, the field can advance toward more effective, scalable, and user-centered solutions for collaborative VR.

\bibliographystyle{ACM-Reference-Format}
\bibliography{CollaborativeEmbodiment.bib}

\end{document}